\documentclass[prb,
superscriptaddress,showpacs,amsmath,amssymb]{revtex4}
\usepackage{amsfonts}
\usepackage{bm}
\usepackage{verbatim}

\usepackage{graphicx}

\begin{document}

\title{Fermionic quartet and vestigial gravity}

\author{G.E.~Volovik}
\affiliation{Low Temperature Laboratory, Aalto University,  P.O. Box 15100, FI-00076 Aalto, Finland}
\affiliation{Landau Institute for Theoretical Physics, acad. Semyonov av., 1a, 142432,
Chernogolovka, Russia}

\date{\today}

\begin{abstract}
We discuss the two-step transitions in superconductors,  where the intermediate state between the Cooper pair state and the normal metal is the 4-fermion condensate, which is called the intertwined vestigial order.
We discuss different types of the vestigial order, which are possible in the spin-triplet superfluid $^3$He, and the topological objects in the vestigial phases. Since in $^3$He the order parameter $A_{\alpha i}$ represents the analog of gravitational tetrads, we suggest that the vestigial states are possible in quantum gravity.  As in superconductors, the fermionic vacuum can experience two consequent phase transitions. At first transition the metric appears as the bilinear combination of tetrads $g_{\mu\nu} =\eta_{ab}< \hat E^a_\mu \hat E^b_\nu>$, while the tetrad order parameter is still absent, $e_\mu^a=< \hat E^a_\mu> =0$. This corresponds to the bosonic Einstein general relativity, which emerges in the fermionic vacuum. The nonzero tetrads $e_\mu^a=< \hat E^a_\mu> \neq 0$  appear at the second transition, where a kind of the Einstein–Cartan–Sciama–Kibble tetrad gravity is formed. This suggests that on the levels of particles, gravity acts with different strength on fermions and bosons. 

\end{abstract}

\maketitle

\section{Introduction. 4-fermion states}
 
Experiments on superfluid $^3$He-B demonstrated the anomalous behaviour, which suggested the possible existence of a new state above the superfluid transition but below the normal state.\cite{Bunkov2000} The possible interpretation was that when the disorder suppresses the anisotropic Cooper pairing, the 4-fermion state with reduced anisotropy can be formed. Such intertwined states are called the vestigial order, see review paper.\cite{Fernandes2019}

There can be many different realizations of the fermionic quartets, see e.g. Eqs. (10.2)-(10.7) in Ref. \cite{Volovik1992} and Eqs. (1) and (2) in Ref. \cite{Volovik1996}. These states have different types of the topological objects.
For example, if the intertwined state has the form $<\Psi\Psi\Psi\Psi>$, the mass of the effective boson is $4m$, which gives rise to the fractional vortices with circulation quantum $2\pi\hbar/4m$. 

An interesting example of the quartet states is the combination of the $S=1$ pairing in the $s$-wave channel and the $S=0$ pairing in the $p$-wave channel, both being forbidden separately.\cite{Balatsky2020}    One may also expect the quartet order parameter, which combines the $p$-wave pairing with ferromagnetism:
\begin{equation}
<\hat a_{k\uparrow} \hat a_{-k\uparrow}  (\hat a^+_{k\downarrow}  \hat a_{k\downarrow} +\hat a^+_{-k\downarrow}  \hat a_{-k\downarrow})> 
\,,
\label{quartet_p}
\end{equation} 
while both the Cooper pairing and ferromagnetism are absent, $<\hat a_{k\uparrow} \hat a_{-k\uparrow}>=0$ and 
$<\hat a^+_{k\uparrow}  \hat a_{k\uparrow} - \hat a^+_{k\downarrow}  \hat a_{k\downarrow} >=0$.
The combination of the $s$-wave equal spin pairing with antiferromagnetism is also possible:
\begin{equation}
<\hat a_{k\uparrow} \hat a_{-k\uparrow}  (\hat a^+_{k\downarrow}  \hat a_{k\downarrow} -\hat a^+_{-k\downarrow}  \hat a_{-k\downarrow})> 
\,.
\label{quartet_p}
\end{equation}  The four-fermion condensate naturally appear in the core of vortices, see Fig. 1(c) in Ref.\cite{SalomaaVolovik}. More on the $4e$ condensates can be found in Refs.\cite{BabaevBook,Tarasewicz2019,Gnezdilov2021,Grinenko2021,Watanabe2022,Fernandes2021,Jian2021,Liu2023a,Liu2023b,Wu2023,Kleinert2008}
  The sextuplets of fermions are possible. In the 2+1 systems they may give the fractional  values ($1/3$ or $1/6$) of the intrinsic Quantum Hall Effect  (without external magnetic field).\cite{Volovik1988}

In Sec. \ref{4nematic} we discuss different types of the quartet vestigial order in the spin-triplet superfluid $^3$He, which correspond to superfluid nematics and normal nematics. We consider the topological objects, which may emerge in these phases. 

Sec. \ref{gravity} is devoted to the extension of the vestigial order to quantum gravity. Here we use the direct analogy between the $^3$He order parameter $A_{\alpha i}$ and the gravitational tetrads, which  suggests the possibility of the vestigial states in quantum gravity.  As in superfluid $^3$He, the fermionic vacuum can experience two consequent phase transitions. At first transition the metric appears as the bilinear combination of tetrads $g_{\mu\nu} =\eta_{ab}< \hat E^a_\mu \hat E^b_\nu>$, while the tetrad order parameter is still absent, $e_\mu^a=< \hat E^a_\mu> =0$. This corresponds to the conventional Einstein general relativity, where the gravitational degrees of freedom are represented by metric. Then the nonzero tetrads $e_\mu^a=< \hat E^a_\mu> \neq 0$  appear at the second transition, where a kind of  the Einstein–Cartan–Sciama–Kibble tetrad gravity develops. 

In Sec. \ref{2bosons} we briefly discuss the boson pairing, which is the analog of the 4-fermion states.

 \section{4-fermion nematics}
 \label{4nematic}
 
Let us consider the vestigial order on example of the spin-triplet $p$-wave superfluid phases of liquid $^3$He. The order parameter in these superfluids is the complex $3\times 3$ matrix $A_{\alpha i}$, where $\alpha$ is the spin vector index, and $i$ is the orbital vector index. This order parameter breaks the symmetry $G$ of the normal state of liquid $^3$He, which contains several symmetry groups: $G=U(1)\times SO(3)_L\times SO(3)_S$. Here $S$ and $L$ denote the spin and orbital rotations, and for simplicity we ignore here the discrete symmetries. 

For the polar phase and for the chiral A-phase the order parameters are:
 \begin{eqnarray}
A_{\alpha i} \,\propto e^{i\Phi} d_\alpha m_i \,\,,\,\, {\rm polar \,\, phase}\,,
\label{Polar}
\\
A_{\alpha i} \,\propto e^{i\Phi} d_\alpha (m_i + in_i)\,\,,\,\, {\rm A-phase}\,.
\label{Aphase}
\end{eqnarray} 
Here  ${\bf d}$ is the unit vector in the spin space; and  ${\bf m}$ and  ${\bf n}$ are the unit vectors in the orbital space with  ${\bf m}\cdot {\bf n}=0$.

 \subsection{Non-superfluid spin nematics}

The many-component order parameter is the typical source of the  intertwined states. That is why It is not excluded there can be different intertwined vestigial states, which separate the polar or the A-phase from the state of the normal liquid.\cite{Volovik1992}
Example is the following  state in which the pair order parameter is absent but the 4-fermionic order parameter $Q_{\alpha\beta}$ is nonzero:
 \begin{equation}
 <A_{\alpha i}>=0\,\,,\,\, Q_{\alpha\beta}=\frac{1}{2} <A_{\alpha i}A^*_{\beta i}+A_{\beta i}A^*_{\alpha i}>  \, \propto d_\alpha d_\beta 
\,,
\label{quartet_nematic}
\end{equation} 
The order parameter $Q_{\alpha\beta}$ is the same as in the nematic liquid crystals. The role of the director in nematic liquid crystals is played by the vector ${\bf d}$ in the spin space.  So, this quartet state represents the spin nematic.
 
In these quartet phases the symmetries, which are broken in the polar or in the A-phase, are partially restored. For example, the polar phase symmetry $SO(2)_L\times SO(2)_S$ is enlarged to the symmetry $U(1)\times SO(3)_L\times SO(2)_S$ in the quartet phase. So there is the following sequence of the symmetry breaking transitions starting with the normal state:
 \begin{eqnarray}
G=U(1)\times SO(3)_L\times SO(3)_S \rightarrow 
\label{SymmetryBreaking1}
\\
\rightarrow H_{\rm nematic}= U(1)\times SO(3)_L\times SO(2)_S \rightarrow
\label{SymmetryBreaking2}
\\
\rightarrow H_{\rm polar}= SO(2)_L\times SO(2)_S
\,.
\label{SymmetryBreaking3}
\end{eqnarray} 
 This intermediate nematic state in Eq.(\ref{SymmetryBreaking2}) is non-superfluid, since the $U(1)$ symmetry is restored.

 \subsection{Superfluid spin nematics}

Another possible vestigial phase is given by:
 \begin{equation}
 <A_{\alpha i}>=0\,\,,\,\, P_{\alpha\beta}=<A_{\alpha i}A_{\beta i}> 
\, \propto e^{2i\Phi}d_\alpha d_\beta  \,.
\label{superfluid_nematic}
\end{equation} 
Here the  symmetry $SO(2)_L\times SO(2)_S$ of the polar phase is extended to $SO(3)_L\times SO(2)_S$ in the quartet phase, and one has the following symmetry breaking scheme:
 \begin{eqnarray}
G=U(1)\times SO(3)_L\times SO(3)_S \rightarrow 
\label{SymmetryBreaking4}
\\
\rightarrow H_{\rm super\,nematic}=SO(3)_L\times SO(2)_S  \rightarrow 
\label{SymmetryBreaking5}
\\
 \rightarrow H_{\rm polar}=SO(2)_L\times SO(2)_S
\,.
\label{SymmetryBreaking6}
\end{eqnarray} 
In intermediate vestigial phase in Eq.(\ref{SymmetryBreaking5}) the $U(1)$ symmetry remains broken, and this state represents the 4-fermion superfluid spin nematic.

  \subsection{Fate of half-quantum vortices in the vestigial states}

 The symmetry breaking scenarios determine the behaviour of the topological defects in transition from the Cooper pairs to quartets. Let us consider this on example of the half-quantum vortex in the polar phase in Eq.(\ref{Polar}). It is the combination of the phase half-vortex with winding $\Phi =\alpha/2$ and the half-disclination  ${\bf d}=\hat{\bf x} \cos\frac{\alpha}{2}+\hat{\bf y} \sin\frac{\alpha}{2}$, where $\alpha$ is the azimuthal angle of cylindrical coordinate system.
 
  In the superfluid nematic quartet phase in Eq.(\ref{superfluid_nematic}) this combined object splits into two separate  topological objects: the isolated half-vortex and the isolated half-disclination.

 \subsection{Fate of monopoles in the vestigial states}

Monopoles experience similar transformations. Let us consider this on example  of the monopole in the planar phase of superfluid $^3$He and in its 4-fermion partners. The order parameter of the planar phase is
 \begin{eqnarray}
A_{\alpha i}\propto e^{i\Phi}\left(R_{\alpha i}- {\hat s}_\alpha  {\hat l}_i \right) \,\,,\,\, {\hat s}_\alpha= R_{\alpha i}  {\hat l}_i \,,
\label{OP1}
\end{eqnarray}
where ${\hat s}_\alpha$ and   ${\hat l}_i$ are spin and orbital unit vectors and $R_{\alpha i}$ is the matrix of rotation, which couples these two vectors.

The corresponding 4-fermion normal spin nematic is given by:
 \begin{equation}
 <A_{\alpha i}>=0\,\,,\,\,Q_{\alpha\beta}= <A_{\alpha i}A^*_{\beta i}> \,
\propto \delta_{\alpha\beta} -s_\alpha s_\beta  
\,,
\label{planar_nematic}
\end{equation} 
while the 4-fermion superfluid spin nematic is given by:
  \begin{equation}
 <A_{\alpha i}>=0\,\,,\,\,P_{\alpha\beta}= <A_{\alpha i}A_{\beta i}> \,
\propto e^{2i\Phi}(\delta_{\alpha\beta} -s_\alpha s_\beta)  
\,.
\label{planar_supernematic}
\end{equation} 

The planar phase in Eq.(\ref{OP1}) contains the point  topological defects -- monopoles. The monopole is the combined object: it is the monopole in spin space  $\hat {\bf s}({\bf r})=\hat {\bf r}$, which is  accompanied by the monopole in the orbital vector, $\hat {\bf l}({\bf r})=\hat{\bf r}$.\cite{Volovik2020} If one tries to split the two monopoles, there appears the analog of the Nambu string which connects the spin and orbital monopoles.\cite{MisirpashaevVolovik1992} 

In the 4-fermion phases (\ref{planar_nematic}) and (\ref{planar_supernematic}), where the symmetry is partially restored, the orbital $\hat {\bf l}$-vector is absent. As a result, the combined monopole in the planar phase, $\hat {\bf s}({\bf r})=\hat {\bf l}({\bf r})=\hat{\bf r}$, transforms to the isolated monopole in the spin nematics vector, $\hat {\bf s}({\bf r})=\hat {\bf r}$. 

In nematics, the monopole with 
$\hat {\bf s}({\bf r})=\hat {\bf r}$ has the same topological charge as the anti-monopole with 
$\hat {\bf s}({\bf r})=-\hat {\bf r}$. This is because the monopole with the topological charge $N=1$ transforms to the anti-monopole with $N=-1$ after circling around the half-disclination.  This is called the influence of the fundamental group $\pi_1$ on
the group $\pi_2$.\cite{VolovikMineev1977}  As a result, in the presence of the half-disclination one has $1+1=0$, which means that two monopoles each with $N=1$ can be annihilated.

One may also construct the orbital 4-fermion  nematic state with the following quartet order parameter:
 \begin{equation}
 <A_{\alpha i}>=0\,\,,\,\,Q_{ij}= <A_{\alpha i}A^*_{\alpha j}> \,  \propto \delta_{ij} -l_i l_j 
\,,
\label{orbital_nematic}
\end{equation} 
and the 4-fermion superfluid orbital nematic:
  \begin{equation}
 <A_{\alpha i}>=0\,\,,\,\,P_{ij}= <A_{\alpha i}A_{\alpha j}>  \, \propto e^{2i\Phi}(\delta_{ij} -l_i l_j)  
\,.
\label{orbital_supernematic}
\end{equation} 
In this symmetry restoration scheme the combined monopole in the planar phase, $\hat {\bf s}({\bf r})=\hat {\bf l}({\bf r})=\hat{\bf r}$, transforms to the isolated orbital monopole with  $\hat {\bf l}({\bf r})=\pm \hat{\bf r}$.

In the two-step transitions one may expect the appearance of the  hybrid defects, composed of two
different types of topological defects with different dimensions. Such combined objects are described by the relative homotopy groups.\cite{VolovikZhang2020}  These include walls bounded by strings and strings terminated by monopoles.\cite{Kibble1982,Makinen2019,Lazarides2023,Eto2023}

  \section{Vestigial gravity}
  \label{gravity}
  
 In some phases of superfluid $^3$He, such as $^3$He-B, the order parameter $A_{\alpha i}$ plays the role of the gravitational triads, which can be extended to tetrads.\cite{Volovik1990,Volovik2022}
The same mechanism of emergent tetrads takes place in the Akama-Diakonov-Wetterich (ADW) gravity,\cite{Akama1978,Wetterich2004,Diakonov2011,Sindoni2012} where tetrads serve as the order parameter, which is the bilinear combinations of fermionic operators (see also Ref.\cite{Vergeles2023}):
 \begin{equation}
 e^a_\mu=<\hat E^a_\mu>\,\,,\,\, 
 \hat E^a_\mu = \frac{1}{2}\left( \Psi^\dagger \gamma^a\partial_\mu  \Psi -  \Psi^\dagger\overleftarrow{\partial_\mu}  \gamma^a\Psi\right) \,.
\label{TetradsFermions}
\end{equation}
The emergent quantum gravity here is of the type of Einstein-Cartan-Sciama-Kibble (ECSK) theory of tetrad gravity, see review in Ref. \cite{Hehl2023}.
The metric, which is the bilinear combination of tetrads, represents the fermionic quartet. 

The analogy with the quartet phases in superfluid $^3$He suggests that the ADW scenario can be extended to incorporate the vestigial states of quantum gravity. These are the states where the bilinear order parameter (tetrad) vanishes,  while the quartet order parameter (metric) is still nonzero:
  \begin{equation}
 <\hat E^a_\mu>=0\,\,,\,\, 
g_{\mu\nu} =\eta_{ab}< \hat E^a_\mu \hat E^b_\nu>\,.
\label{GR}
\end{equation}
 The  vestigial order in Eq.(\ref{GR}) describes the emergence of the Einstein general relativity in terms of metric fields $g_{\mu\nu}$. The metric allows to write the Einstein action and the action for bosonic matter, such as Maxwell action for the bosonic gauge fields.  So, this symmetry breaking gives rise to the gravity for bosons, although it emerges in the fermionic vacuum. The metric here is the analog of the order parameter in biaxial nematics in the orbital space.

The further spontaneous symmetry breaking is the breaking of the spin rotation symmetry by the tetrad order parameter $e^a_\mu$ in Eq.(\ref{TetradsFermions}). This gives rise to  the Weyl-Dirac action for fermions
and to the Einstein-Cartan-Sciama-Kibble (ECSK) tetrad gravity, which interacts  also with fermions. The sequence of the symmetry breaking phase transitions is now: 
 \begin{equation}
{\rm disorder} \rightarrow \,{\rm GR} \, \rightarrow {\rm ECSK}\,.
\label{GravityScheme}
\end{equation}

Due to the quartic correlators, the ECSK gravity may have the memory on  the  vestigial gravity:
\begin{eqnarray}
g_{\mu\nu} =\eta_{ab}e^a_\mu e^b_\nu+ \tilde g_{\mu\nu} \,,
 \label{TetradsMetric1}
 \\
 \tilde g_{\mu\nu} =\eta_{ab}(< \hat E^a_\mu \hat E^b_\nu> - < \hat E^a_\mu>< \hat E^b_\nu>)\,.
\label{TetradsMetric2}
\end{eqnarray}
This means that on the levels of particles, gravity acts with different strength on fermions and bosons. Fermions interact with tetradic part of the metric, while bosons interact with the full metric. Thus the Equivalence Principle can be violated, i.e.  a boson and a fermion in a given gravitational field do not follow the same trajectories.

Another example of the  vestigial gravity can be constructed using the gravity with varying signature of metric:\cite{BondarenkoZubkov2022}
 \begin{equation}
 <\hat E^a_\mu>=0\,\,,\,\, 
g_{\mu\nu} =< Q_{ab}\hat E^a_\mu \hat E^b_\nu>\,.
\label{BondarenkoZubkov}
\end{equation}
Here $Q_{ab}$ is the dynamical field, which in particular determines the signature of the metric.

  \section{2-boson superfluid}
  \label{2bosons}
 
The quartet of fermions is equivalent to bosonic pair. 
It is natural that in the presence of the Bose condensate $<\hat a_0>=\sqrt{n_0} \neq 0$ one has the nonzero value of the pair condensate:
\begin{equation}
\chi = \sum_{k\neq 0}  <\hat a_k \hat a_{-k} >  \neq 0\,.
\label{chi}
\end{equation} 
 This can be seen from the simplest 4-boson model of interaction:
\begin{eqnarray}
H = V_0 a_0a_0a_0^+ a_0^+ + V_1 \sum_{k\neq 0} ( \hat a_0^+ \hat a_0^+ \hat a_k \hat a_{-k} + \hat a_0 \hat a_0 \hat a_k^+ \hat a_{-k}^+) +
\nonumber
\\
+V_2 \sum_{k\neq 0}  \hat a_k \hat a_{-k} \sum_{p\neq 0}  \hat a_p^+ \hat a_{-p}^+
\,.
\label{H}
\end{eqnarray} 
From Eq.(\ref{H}) one obtains the energy of the $\chi$-field:
 \begin{equation}
E(\chi)= V_1n_0 (\chi + \chi^*) + V_2 |\chi|^2
\,,
\label{F}
\end{equation} 
and minimization gives 
\begin{equation}
\chi = - n_0 \frac{V_1}{V_2}  \,.
\label{chi_equil}
\end{equation} 
This is valid for $V_1\ll V_2$. But in general it is not excluded that the pairing can be dominating, if the Bose condensate is highly suppressed by interaction. This in principle, may happen in liquid $^4$He, where the condensate is highly depleted. In this case the singly quantized vortex splits into two half-quantum vortices, bounded by Kibble wall.\cite{Volovik2002}

Moreover, it is not excluded that in a very strong interaction, the 2-boson pairing occurs first, 
$< \hat a_k \hat a_{-k}> \neq 0$,  while the Bose condensate is still absent, $<\hat a_0> = 0$.\cite{Nozieres1982,Schmidt2006} Such pairing may exist just above the transition to the Bose condensate.

On pair condensate see also \cite{Pashitskii2002,Levin2023}.
Pair condensate inside vortices can be found in Refs.\cite{SalomaaVolovik,Takeuchi2021}.

  \section{Discussion}
   \label{discussion}

The unexpected consequence of the  vestigial order states in superfluids and superconductors is the possibility of the similar states in quantum gravity. This suggests that the traditional Einstein gravity in terms of metric  and the Einstein-Cartan gravity in terms of tetrads are separated by the symmetry breaking phase transition. If so, this may have consequences for cosmology. For example, the state with Einstein gravity and the state with the Einstein-Cartan gravity may occupy different regions of spacetime. On the levels of particles, gravity acts with different strength on fermions and bosons. The Equivalence Principle does not hold, i.e.  a boson and a fermion in a given gravitational field do not follow the same trajectories. This may also influence cosmology.\cite{Barrow2004}

 \end{document}